\documentclass[aps,prl,reprint,balancelastpage,nofootinbib,preprintnumbers,superscriptaddress]{revtex4-1}

\usepackage[english]{babel}
\usepackage{latexsym,amssymb,amsmath,amsfonts,physics,url,mathrsfs,dsfont}
\usepackage{graphicx}
\usepackage{comment}
\usepackage[table]{xcolor}
\usepackage{orcidlink}
\usepackage{tikz}
\usepackage[caption=false,labelformat=empty]{subfig}
\usepackage{bm}
\usepackage{upgreek}
\usepackage{amsfonts}
\usepackage{diagbox}

\newcommand{\sv}[2]{\hat{#1}_{\langle #2 \rangle}}

\def\DD{\mathsf{D}_\ell} 
\def\Dd{\mathsf{d}_\ell} 

\def\RS{R_\text{S}}
\def\Rb{R_{\rm b}}

\def\ZC{\mathbf{L}}

\def\nn{\nonumber}
\def\sch{Schwarzschild }

\usepackage{cleveref}
\crefname{equation}{Eq.}{Eqs.}
\crefname{section}{Sec.}{Secs.}
\crefname{table}{Tab.}{Tabs.}
\crefname{figure}{Fig.}{Figs.}

\begin{document}

\begin{flushleft}
    \hfill CERN-TH-2025-247
\end{flushleft}

\title[]{Dynamical Love Numbers for Black Holes and Beyond \\
from Shell Effective Field Theory}

\author{D. Kosmopoulos\orcidlink{0000-0001-9976-3435}}
\affiliation{Department of Theoretical Physics, CERN, CH-1211 Geneva 23, Switzerland}

\author{D. Perrone\orcidlink{0000-0003-4430-4914}}
\affiliation{Department of Theoretical Physics and Gravitational Wave Science Center,  \\
24 quai E. Ansermet, CH-1211 Geneva 4, Switzerland}

\author{M. Solon\orcidlink{0000-0003-1445-6888}}
\affiliation{Mani L. Bhaumik Institute for Theoretical Physics, Department of Physics and Astronomy, University of California Los Angeles, Los Angeles, CA 90095, USA}


\begin{abstract}
\noindent
We construct a novel effective field theory for a compact body coupled to gravity, whose key feature is that the dynamics of gravitational perturbations is explicitly determined by known solutions in black hole perturbation theory in four dimensions. In this way, the physics of gravitational perturbations in curved space are already encoded in the effective field theory, thus bypassing the need for the higher-order calculations that constitute a major hurdle in standard approaches. Concretely, we model the compact body as a spherical shell, whose finite size regulates short-distance divergences in four dimensions and whose tidal responses are described by higher-dimensional operators. As an application, we consider scalar perturbations and derive new results for scalar Love numbers through ${\cal O} (G^9)$ for Schwarzschild black holes and for  generic compact bodies. Finally, our analysis reveals an intriguing structure of the scalar black-hole Love numbers in terms of the Riemann zeta function, which we conjecture to hold to all orders.
\end{abstract}
\maketitle

\section{Introduction}

The rapid growth of gravitational-wave astronomy in recent years has driven the development of new field-theoretic approaches for gravitational-wave physics (see, e.g.,~\cite{Buonanno:2022pgc,Adamo:2022dcm,Bjerrum-Bohr:2022blt,Kosower:2022yvp} for reviews). 
While a large focus of this program is the two-body problem~\cite{Cheung:2018wkq,Bern:2019nnu,Bern:2021dqo,Dlapa:2021vgp,Dlapa:2021npj,Dlapa:2022lmu,Kalin:2022hph,Damgaard:2023ttc,Jakobsen:2023pvx,Jakobsen:2023ndj,Dlapa:2024cje,Driesse:2024xad,Bern:2024adl,Dlapa:2025biy,Bern:2025zno}, the same tools have now been applied for understanding properties of a single compact body, such as tidal deformability. 
The tidal response of a body to an external gravitational field is characterized by Love numbers~\cite{Love1909}, and plays an essential role in determining its internal structure. Probing the nature of astrophysical compact bodies, such as black holes and neutron stars, by measuring the imprints of Love numbers on the gravitational waveform is one of the principal scientific goals of gravitational-wave astronomy (see, e.g.,~\cite{Flanagan:2007ix,ET:2025xjr}).

From the point of view of effective field theory (EFT), Love numbers are related to Wilson coefficients of tidal operators in the worldline effective action~\cite{Goldberger:2004jt,Goldberger:2005cd,Porto:2005ac,Goldberger:2006bd,Porto:2008jj,Porto:2008tb,Porto:2010tr,Porto:2012as,Levi:2014gsa,Levi:2015ixa,Levi:2019kgk,Mogull:2020sak,Kalin:2020lmz,Levi:2020lfn,Jakobsen:2021zvh,Liu:2021zxr,Creci:2021rkz,Steinhoff:2021dsn,Kim:2021rfj,Cho:2022syn,Mandal:2022ufb,Mandal:2022nty,Kim:2022bwv,Kim:2022pou,Levi:2022rrq,Levi:2022dqm,Edison:2023qvg,Creci:2023cfx,Mandal:2023hqa,Mandal:2023lgy,Creci:2024wfu,Mandal:2024iug,Saketh:2024juq,Edison:2024owb,Hoogeveen:2025tew,Bhattacharyya:2025slf} (or in the quantum-field-theory action~\cite{Brandhuber:2019qpg,Haddad:2020que,Aoude:2020ygw,AccettulliHuber:2020dal,Cheung:2020sdj,Cheung:2020gbf,Bern:2020uwk,Aoude:2020ygw,Aoude:2020onz,Kosmopoulos:2021zoq,Chiodaroli:2021eug,Aoude:2022thd,Aoude:2022trd,Bern:2022kto,Cangemi:2022bew,Aoude:2023vdk,Haddad:2023ylx,Bern:2023ity,Cangemi:2023bpe,Alaverdian:2024spu,Akpinar:2024meg,Bohnenblust:2024hkw,Correia:2024jgr,Aoude:2025xxq,Alaverdian:2025jtw,Akpinar:2025bkt,Akpinar:2025byi}), which systematically parametrize finite-size corrections to the point-particle limit. 
This provides a gauge-invariant definition of Love numbers, avoiding coordinate-dependent ambiguities (see, e.g.,~\cite{Pani:2015nua,Pani:2015hfa,Gralla:2017djj}).
Static Love numbers of black holes vanish at both the linear~\cite{Fang:2005qq,Damour:2009vw,Binnington:2009bb,Damour:2009va,Kol:2011vg,Gurlebeck:2015xpa} and nonlinear order~\cite{Combaluzier-Szteinsznaider:2024sgb,Kehagias:2024rtz} (see also \cite{Hui:2020xxx,LeTiec:2020spy,Charalambous:2021mea,Hui:2022vbh,Ivanov:2022qqt,Ivanov:2022hlo,Kehagias:2022ndy,Saketh:2023bul,Perry:2023wmm,Chakraborty:2023zed,Riva:2023rcm,DeLuca:2023mio,Gounis:2024hcm,Kehagias:2024yzn,DeLuca:2024ufn}), so a nonzero measurement would signal physics beyond general relativity (GR)~\cite{Cardoso:2017cfl}. 
This vanishing has also been the subject of recent studies as a test bed for our understanding of symmetry and naturalness~\cite{Porto:2016zng,Charalambous:2021kcz,Hui:2021vcv,Charalambous:2022rre,BenAchour:2022uqo,Rai:2024lho,Sharma:2024hlz,Sharma:2025xii,Berens:2025jfs,Lupsasca:2025pnt,Charalambous:2025ekl,Parra-Martinez:2025bcu}. 
On the other hand, dynamical Love numbers are generally nonzero and can be determined by matching observables in the worldline EFT and in GR~\cite{Barack:2023oqp,Bini:2024icd,Ivanov:2024sds,Akhtar:2025nmt} (see also~\cite{Bautista:2021wfy,Bonelli:2021uvf,Consoli:2022eey,Bautista:2022wjf,Bautista:2023sdf}). 
This matching requires high-order calculations which are quite challenging with traditional methods (see, however, recent progress in~\cite{Caron-Huot:2025tlq,Combaluzier--Szteinsznaider:2025eoc,Kobayashi:2025vgl,Chakraborty:2025wvs}).

\begin{figure}[t]
    \centering
    \includegraphics[width=\linewidth]{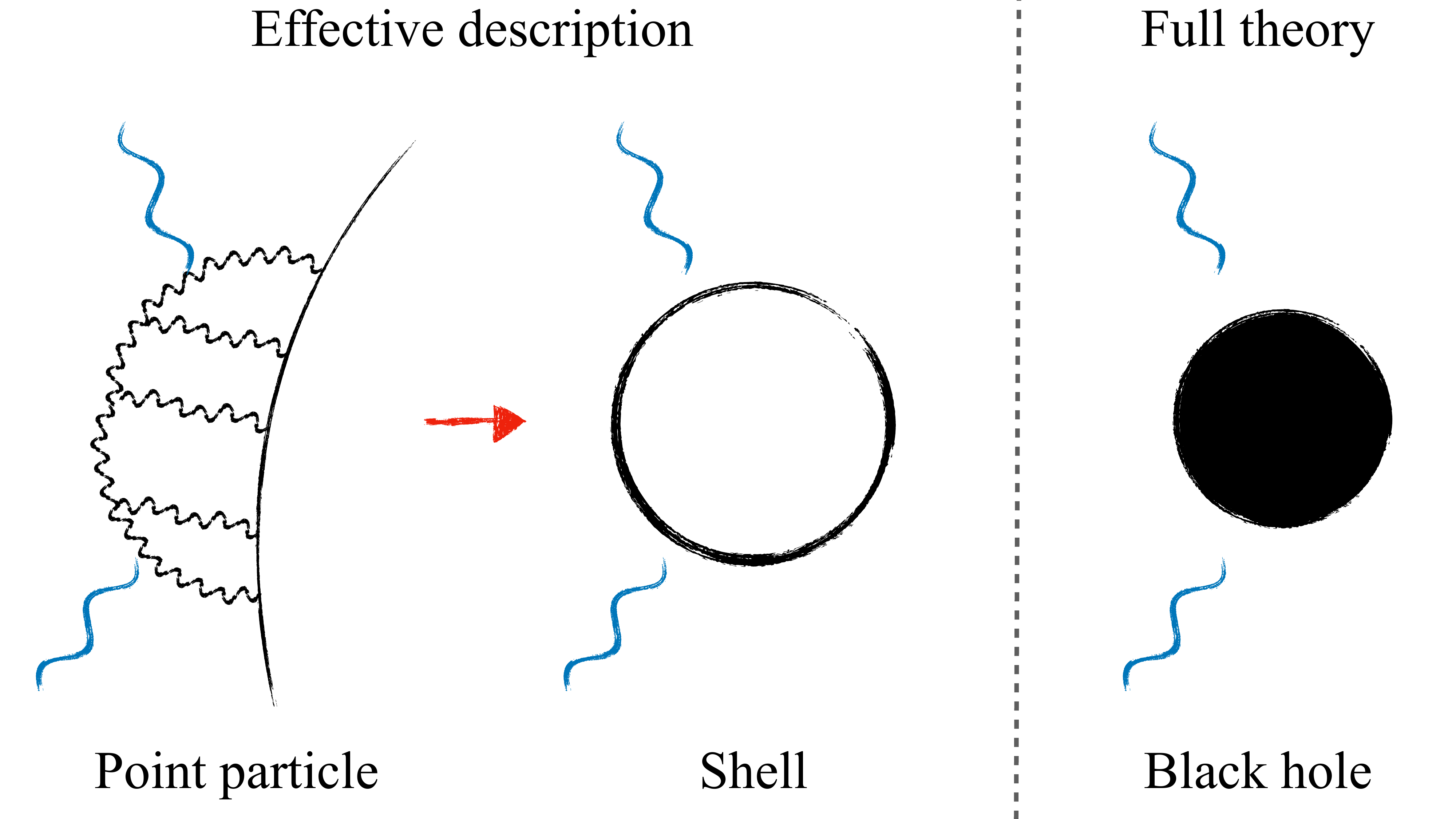}
    \caption{
    In the standard point-particle approach, calculating the tidal response involves loop contributions.
    By replacing the point particle with a shell, our EFT framework naturally incorporates known BHPT solutions in four dimensions, thus bypassing the need for higher-order integration. Moreover, the matching to full theory is simplified since gravitational perturbations are described by the same solutions.
    }
    \label{fig:shell}
\end{figure}

In this Letter, we address this problem by developing a new EFT for a compact body coupled to gravity, which we call Shell EFT. This approach directly leverages known solutions in black hole perturbation theory (BHPT) in four dimensions, and thus avoids the need for loop diagrams that build up effects due to the black-hole spacetime (see \cref{fig:shell}). In essence, Shell EFT employs a synergy of tools from quantum field theory and GR, and offers a promising avenue for achieving the theoretical precession necessary for future gravitational-wave experiments.

As a first application, we consider scalar perturbations and derive new results for scalar Love numbers of generic compact objects through ${\cal O}(G^9)$. For the specific case of black holes, we agree with previous results through ${\cal O}(G^7)$~\cite{Caron-Huot:2025tlq}. Moreover, our expression exhibits interesting structure in terms of the Riemann zeta function, allowing us to conjecture results to all orders.

\section{Shell Effective Field Theory}

The basic idea of Shell EFT is to describe a generic compact body as a spherical shell so that its finite radius $R$ acts as a regulator for short-distance divergences in four dimensions. This allows us to solve for the dynamics of gravitational perturbations in terms of known solutions in BHPT in four dimensions~\cite{Leaver:1986gd,Leaver:1986vnb,Mano:1996vt}. 

The above ingredients are reflected in the background metric, which is flat inside the shell and Schwarzschild outside:
\begin{align}
    \dd s_+^2 &= f(r) \dd t^2 - f(r)^{-1}\dd r^2 - r^2\dd \Omega^2\,, \quad r>R\,,
    \nonumber\\ 
    \dd s^2_{-}&= \dd \tilde t\hspace{.7pt}^2 - \dd \tilde r^2 - \tilde r^2\dd \Omega^2\,, \quad \tilde r<R\,,
    \label{eq:metric}
\end{align}
where $f(r) = 1-\RS/r$ with $\RS = 2GM$ and $R> \RS$. The junction conditions 
\begin{align}
    \tilde t = t\sqrt{f(R)}\,,
    \quad
    \tilde r= R + \frac{r-R}{\sqrt{f(R)}}\,,
    \label{eq:junction}
\end{align}
ensure continuity of the metric on the shell $r=\tilde r =R$~\cite{Israel:1966rt}.\footnote{
Thin shells have found many applications in GR, e.g., in modeling gravitational collapse~\cite{Oppenheimer:1939ue,Israel:1967zz}, 
in the construction of traversable wormholes~\cite{Visser:1989kh}, 
in the description of vacuum–decay bubbles and cosmological phase transitions~\cite{Coleman:1980aw},
as well as in the development of the membrane paradigm for black holes~\cite{MacDonald:1982zz,Thorne:1986iy}. They have also recently appeared in the study of scattering amplitudes~\cite{Bjerrum-Bohr:2025lpw}.}

For the application to Love numbers, we consider the scattering of a scalar wave $\Phi$ with frequency $\omega \ll 1/R$ off the shell, and model the tidal responses of the shell using higher-dimensional operators. These are captured by the action
\begin{align}
    S_{\rm Shell}=    \int \frac{\dd^3y \sqrt{\gamma}}{4\pi R^2} \sum_{\ell,n}\frac{\hat c_{\ell,n}}{\ell!} \varphi (u^{\mu} \nabla_{\mu})^n \DD \phi \,,
    \label{eq:shellAction}
\end{align}
while the bulk dynamics of the scalar field are described by the Klein-Gordon action,
\begin{align}\label{eq:KG}
    S_{\rm Bulk} = \int\dd^4 x \sqrt{-g} \, g^{\mu\nu} \partial_{\mu}\varphi \partial_{\nu}\phi \, .
\end{align}
We describe the elements entering \cref{eq:shellAction,eq:KG} in turn.

In \cref{eq:shellAction}, we omit the terms that source the background geometry. These are fixed by the junction conditions and do not enter our calculation.

In order to account for dissipative tidal effects, we adopt the \textit{in-in} or \textit{Schwinger-Keldysh} formalism~\cite{Schwinger:1960qe,Keldysh:1964ud} where the field $\Phi$ is doubled: $\Phi \rightarrow (\varphi, \phi)$. 
The field $\phi$ represents the classical field that we solve for while $\varphi$ represents the fluctuations that source it.

We take the shell coordinates to be $y^a = (t,\uptheta,\upphi)$, such that the metric induced on the shell's surface is given by
\begin{align}
    \gamma_{ab} = g_{\mu\nu} e_a^\mu e_b^\nu \Big\rvert_{r=R} = 
    \begin{pmatrix}
        f(R) & 0 & 0 \\
        0 & -R^2 & 0\\
        0 & 0 & -R^2 \sin^2 \uptheta
    \end{pmatrix} \,,
\end{align}
where $e_a^\mu = \partial x^\mu/\partial y^a$.

The tidal responses of the shell are built out of $e_a^\mu \nabla_\mu$ as an expansion in the size of the compact body over the wavelength of the perturbation.
Each operator is associated with a free Wilson coefficient $\hat c_{\ell,n}$, which is determined by matching to full theory. 

Since Lorentz-boost symmetry is broken, time derivatives $u^{\mu} \nabla_{\mu}$ are included separately, and we identify the local velocity field of the shell as $u^\mu \equiv e_0^\mu$. 
Given that the compact body preserves rotational symmetry, angular derivatives are given by the rotationally invariant combination
\begin{align}
    \DD = \frac{(-1)^\ell (2\ell+1)}{2^\ell f(R)^\ell R^{2\ell}} \prod_{j=0}^{\ell-1} \left(\nabla_\Omega^2 +j(j+1) \right) \,,
    \label{eq:DlDef}
\end{align}
which is conveniently chosen to project out all angular momenta $\ell'<\ell$.

Ultimately, we are interested in Love numbers defined in the point-particle limit, and we consider a minimal Shell-EFT construction for this purpose.
With this goal in mind, the class of operators in \cref{eq:shellAction} forms a complete set.
We may see this in two ways:
First, as is common practice in the study of EFTs, we establish that we have a complete basis of operators by evaluating them \textit{on shell}. 
By looking at the effect of this class of operators on the equation of motion (EoM), we see that we saturate the required freedom (see comments below \cref{eq:Love_numbers}).
Second, in the supplemental material \cite{suppmat}, we demonstrate the one-to-one correspondence between our effective action and the worldline EFT.
Since the above operators form a complete set, we do not include operators with radial derivatives or curvature tensors. As we discuss below, radial derivatives are in fact discontinuous on the shell.

\section{Dynamics in the Shell EFT}

We proceed to solve the EoM for the classical scalar field $\phi$ in Shell EFT,
\begin{align}
    \frac{\delta S_{\rm Bulk}}{\delta \varphi} + \frac{\delta S_{\rm Shell}}{\delta \varphi} = 0\,.
    \label{eq:eom}
\end{align}
For obtaining the Wilson coefficients of $S_{\rm Shell}$, it is sufficient to consider a static shell, hence the dynamics are fully captured by \cref{eq:eom}. 
We write the solution as 
\begin{align}
    \phi = \phi^+ \Theta(r-R) + \phi^- \Theta(R-r)\,,
    \label{eq:inOutDecomposition}
\end{align}
and solve separately for the dynamics inside and outside the shell, as well as the contribution at $r=R$, which involves the Wilson coefficients.

\subsection{Outside solution}

Using the spherical symmetry of the \sch background we can write the scalar field as a partial-wave expansion, 
\begin{equation}
    \phi^+=e^{-i\omega t}\sum_{\ell} \psi^{+}_{\ell}(r)P_{\ell}(\cos \uptheta)\,, \quad r>R\,,
    \label{eq:ansatz-out}
\end{equation}
and the EoM as
\begin{equation}
    \left(\frac{1}{r^2}\partial_r r^2 f(r)\partial_r + \frac{\ell(\ell+1)}{r^2}+ \frac{\omega^2}{f(r)}\right)\psi_{\ell}^+(r)=0\,.
\end{equation}
The solutions to this equation are known~\cite{Leaver:1986gd,Leaver:1986vnb,Mano:1996vt} and can be written as
\begin{align}
    \psi_{\ell}^+=\frac{1}{z}
        \sum_{n=-\infty}^{+\infty} b_{n}^{\ell} \Big( A_{\ell} F_{\nu_{\ell}+n} + (-1)^n B_{\ell} F_{-\nu_{\ell}-n-1}\Big)\,,
    \label{eq:sol-out}
\end{align}
where we defined $z=\omega (r - \RS)$, and the coefficients $b_{n}^{\ell}$ and the \textit{renormalized angular momentum} $\nu_{\ell}$ are found recursively (see \cite{Castro:2013kea,Castro:2013lba,Bonelli:2021uvf,Bonelli:2022ten,Bautista:2023sdf,Nasipak:2024icb,Ivanov:2025ozg} for interpretations of $\nu_\ell$). 
The boundary coefficients $A_{\ell}$ and $B_{\ell}$ characterize the compact body and are determined by matching to full theory. The Coulomb wavefunctions $F_{j}(z)$ are defined as
\begin{align}
    F_{j}(z) = \,&\frac{\Gamma(j+1+\eta )}{\Gamma(2j+2)} 2^j z^{j+1}e^{-i z}\,\times \nn\\
    &M(j+1+\eta, 2j+2, 2i z)\,, \quad \eta= i\omega\RS\,, 
\end{align}
where $M(a,b,z)$ is the confluent hypergeometric function,
\begin{equation}
    M(a,b,z)=\sum_{s=0}^{\infty}\frac{(a)_s}{(b)_s s!}z^s\,, \quad (a)_s= \frac{\Gamma(a+s)}{\Gamma(a)}\,.
\end{equation}

\subsection{Inside solution}

In the interior of the shell, the spacetime is flat and the wave equation can be solved in terms of Bessel functions. 
We have 
\begin{equation}
    \phi^- = e^{-i\tilde \omega \tilde t} \sum_{\ell} \psi_{\ell}^{-}(\tilde r)P_{\ell}(\cos\uptheta)\,, \quad \tilde r < R\,,
\end{equation}
with
\begin{equation}
    \psi_{\ell}^-(\tilde r ) = \tilde A_{\ell} j_{\ell}(\tilde \omega \tilde r)\,,
\end{equation}
where we only keep the solution that is regular at the origin. 
Demanding continuity of the solution across the shell yields two conditions, 
\begin{align}
    \psi_{\ell}^+(R) = \psi_{\ell}^-(R) \equiv \psi_{\ell}(R)\,, \quad f(R)\,\tilde \omega^2 = \omega^2 \,,
    \label{eq:continuity}
\end{align}
which allow us to specify the two elements that characterize the inside solution, i.e., the coefficient $\tilde A_{\ell}$ and the frequency $\tilde \omega$.
Continuity of the solution is a requirement for the EoM to be well defined; it guarantees that no terms worse than $\delta(r-R)$ appear despite the piecewise form of the solution in \cref{eq:inOutDecomposition} and the EoM in \cref{eq:eom} being a second-order differential equation.
The junction conditions for the metric in \cref{eq:junction} follow in the same way~\cite{Israel:1966rt}.

\subsection{Dynamics across the shell}

At $r=R$, both $S_{\rm Bulk}$ and $S_{\rm Shell}$ contribute. The contribution from the bulk is
\begin{align}
    \frac{\delta S_{\rm Bulk}}{\delta \varphi} &= e^{-i\omega t} \delta(r-R)f(R) \times \nn \\ 
    &\sum_{\ell} P_{\ell}(\cos\uptheta)\left(\partial_{ r}\psi_{\ell}^+ - \partial_r\psi_{\ell}^-\right)\,,
    \label{eq:dS_bulk}
\end{align}
while the shell contribution is
\begin{align}
    \frac{\delta S_{\rm Shell}}{\delta \varphi}&=e^{-i\omega t}\frac{\delta(r-R)\sqrt{f(R)}}{4\pi R^2} \times \nn \\&\sum_{\ell} P_{\ell}(\cos \uptheta) \frac{(2\ell+1)!}{\ell!\, 2^{\ell}R^{2\ell} f(R)^{\ell}}  \psi_{\ell}(R) \, \mathcal{F}_{\ell} \,,
    \label{eq:dS_shell}
\end{align}
where we defined
\begin{align}
    \mathcal{F}_{\ell} = \sum_{n}(-i\omega)^n c_{\ell,n}\,, 
    \label{eq:wilson_coeff}
\end{align}
and
\begin{align}
       c_{\ell,n} = \sum_{\ell' \leq \ell} 
        \frac{2^{\ell}\, f(R)^{\ell}\, R^{2\ell}}{2^{\ell'} f(R)^{\ell'} R^{2\ell'}} 
        \frac{\ell!}{\ell'!}
        \frac{(2\ell'+1)}{(2\ell+1)!}
        \frac{E_{\ell' \ell}}{(-1)^{\ell'}} \hat c_{\ell',n} \,,
        \label{eq:rotation}
\end{align}
in terms of 
\begin{align}
\label{eq:E_projector}
    E_{\ell' \ell} = \prod_{j=0}^{\ell'-1} \left(-\ell(\ell+1) +j(j+1) \right) \,.
\end{align}
We may think of \cref{eq:rotation} as a rotation between the coefficients $\hat c_{\ell,n}$ that appear in the Lagrangian and the coefficients $c_{\ell,n}$ that parametrize the partial-wave expansion of the EoM (i.e., the on-shell coefficients). The normalization is such that $c_{\ell,n} \to \hat c_{\ell,n}$ as $R \to 0$ if there were no divergences associated with that limit.

Combining \cref{eq:dS_bulk,eq:dS_shell}, we arrive at one of the main results of this paper,
\begin{equation}
    \mathcal{F}_{\ell} = 4\pi \frac{\ell! \,2^{\ell}\sqrt{f(R)}^{2\ell+1}R^{2\ell+2}}{(2\ell+1)!}\frac{\partial_{r}\psi_{\ell}^- - \partial_r\psi_{\ell}^+}{\psi_{\ell}}\Big|_{R}\,.
    \label{eq:Love_numbers}
\end{equation}
This is an exact solution for the Wilson coefficients of our Shell EFT in terms of the coefficients $A_\ell$ and $B_\ell$ in $\psi_\ell^+$.
This equation yields one relation per Wilson coefficient $c_{\ell,n}$, which justifies our claim that \cref{eq:shellAction} contains a complete set of operators.

\section{Love Numbers}

We determine Love numbers from the result in~\cref{eq:Love_numbers} by matching to full theory and then taking the point-particle limit. In full theory, we consider a compact body with radius $\Rb$, and the outside solution $\psi_\ell^+{}^{\rm full}$ takes the form specified in~\cref{eq:ansatz-out,eq:sol-out} but with boundary coefficients $A_\ell^{\rm full}$ and $B_\ell^{\rm full}$. 
We match by taking $R>\Rb$ and demanding that the outside solutions in Shell EFT and full theory are equal for $r>R$, which implies   
$A_\ell = A_\ell^{\rm full}$ and $B_\ell = B_\ell^{\rm full}$. Thus, matching simply amounts to using the coefficients $A_\ell^{\rm full}$ and $B_\ell^{\rm full}$ in~\cref{eq:Love_numbers}, and we see directly how different boundary conditions in full theory translate to different Wilson coefficients in Shell EFT. In particular, for black holes, the boundary coefficients $A_\ell^{\rm full}$ and $B_\ell^{\rm full}$ impose the purely incoming condition at the horizon.

We identify $c_{\ell,n}$ from \cref{eq:Love_numbers} by expanding in $\omega$ and picking the appropriate term.
To make contact with the point-particle description of compact objects, we further expand in $\RS$, effectively shrinking the horizon to zero size and manifesting the perturbative or loop expansion, and then take the $R\rightarrow 0$ limit,
which is well defined after the expansion in $\RS$.
Following this procedure, the resulting Wilson coefficients take the form,
\begin{align}
    c_{\ell,n} = \lambda_{\ell,n} + \mathcal{O}\left(\RS/R\right)  \,,
\end{align}
where $\lambda_{\ell,n}$ depends on $R$ only through powers of $\log(\RS/R)$ for black holes or $\log(\Rb/R)$ for a compact body of radius $\Rb$.
This simple EFT requirement provides a highly nontrivial check of the computation, as it emerges from powers of $\log(\omega \RS)$ or $\log(\omega \Rb)$ in $A^{\rm full}_\ell$ and $B^{\rm full}_\ell$ combining with powers of $\log(\omega R)$ in the Coulomb wavefunctions (see \cref{eq:sol-out}).
Finally, we identify the Love numbers for the compact object as
\begin{equation}
    \bar{\mathcal F_\ell} \equiv \sum_{n}(-i\omega)^n \lambda_{\ell,n} \,.
    \label{eq:fbar}
\end{equation}

As is true for Wilson coefficients in general, Love numbers are computed using a particular regularization scheme, e.g., the radius $R$ in Shell EFT or dimensional regularization. We discuss how to identify the scheme-independent terms below~\cref{eq:universal} and in the supplemental material~\cite{suppmat}.
We note that the outside solution $\psi^+_{\ell}$ contributes to both scheme-dependent and scheme-independent terms, while $\psi_\ell^-$ contributes only to scheme-dependent ones.

For black holes, our results through ${\cal O}(G^9)$ verify the vanishing of static Love numbers~\cite{Fang:2005qq,Damour:2009vw,Binnington:2009bb,Damour:2009va,Kol:2011vg,Gurlebeck:2015xpa} as well as the scheme-independent parts in recent lower-order results for dynamical Love numbers, which were obtained 
using dimensional regularization~\cite{Ivanov:2024sds,Caron-Huot:2025tlq}.
Our complete expressions for the Love numbers of black holes are given in the supplemental material \cite{suppmat}.
Our results for the Wilson coefficients for generic compact objects (in terms of arbitrary coefficients $A_\ell^{\rm full}$ and $B_\ell^{\rm full}$) and for black holes are presented in the ancillary files of the supplemental material \cite{suppmat} in terms of ${\mathcal{F}}_{\ell}$.

\section{All-Orders Structure and Resummation}

Our analysis through $\mathcal{O}\left(G^9\right)$ reveals a deep structure for the black-hole Love numbers in terms of the Riemann zeta function.
By extrapolating this structure to all orders in $\RS$ and resumming the perturbative expression, we see hints of the black-hole quasi-normal-mode (QNM) spectrum. 

We conjecture that the expansion coefficients $\lambda_{\ell,n}$ of the dynamical black-hole Love numbers $\bar{\mathcal F_\ell}$ for all $\ell$ and $n \geq 0$ may be written in the following form:
\begin{align}
    \frac{(-1)^n}{4 \pi \RS^{2\ell+n+1}} \lambda_{\ell,n} &= \ZC_\ell \, \zeta_{n-1} + \sum_{\bm a} \big(\ZC_{\ell,n}^{(2)}\big)_{a_1,a_2} \zeta_{a_1} \, \zeta_{a_2} \\
    &+ \sum_{\bm a} \big(\ZC_{\ell,n}^{(3)}\big)_{a_1,a_2,a_3} \zeta_{a_1}\, \zeta_{a_2}\, \zeta_{a_3} + \ldots \,, \nn
    \label{eq:zeta}
\end{align}
where the summations are over the domains
\begin{align}
    0 \leq a_{i} \leq n-N\,, 
    \qquad \sum_{i=1}^N a_i = n-N \,,
\end{align}
with $N$ being the number of indices.
The coefficient $\lambda_{\ell,n}$ is organized in terms of $n$ sums of decreasing transcendental weight.\footnote{The transcendental weight $w$ satisfies $w[q]=0$ for any rational number $q$, $w[\pi] = 1$, $w[\zeta_n] = n$ for any non-negative integer $n$, $w[\log(x)] = 1$ for any $x>0$, and $w[xy] = w[x]+w[y]$.}
We identify $\zeta_1 \equiv -\log\left(\RS/R\right)/2$ as a regularized value for $\zeta_1$, a choice that is natural from the standpoint of transcendental weight.
Our conjecture incorporates the vanishing of the static Love numbers for $n=0$, since $\lambda_{\ell,0}$ is given in terms of zero sums.
The $\ZC_{\ell,n}^{(N)}$ are matrices of  rational numbers which are symmetric under the interchange of any two indices $a_i$, and the scheme-independent subset of their entries depends on $N$ and $\ell$ but not $n$ or $\bm a$,
\begin{align}
    \left(\ZC^{(N)}_{\ell,n}\right)_{\bm a} = \ZC^{(N)}_{\ell} \,, \quad 
    \text{if at most one $a_i$ is zero}
     \,.
     \label{eq:universal}
\end{align}
Only entries with $a_i=0$ for at least two values of $i$ are scheme dependent: For a term of the form $A \equiv \zeta_0^2 Z$, where $Z$ is a product of zetas, there exist a term $B \equiv \zeta_1 Z = -\log\left( \RS/R\right)Z/2$; by shifting the argument of the logarithm in $B$, we can change the value of $A$.\footnote{We only allow for shifts that produce terms of zero transcendental weight. See also the supplemental material \cite{suppmat}.}
We note that there is freedom in choosing $\ZC$ in \cref{eq:zeta} due to relations of the form $\zeta_2^2 \propto \zeta_4$; this freedom is crucial for observing the universality of \cref{eq:universal} and performing the resummation we discuss next.

Although observed in Shell EFT, this structure pertains to the scheme-independent parts of the Love numbers and should arise in any scheme, e.g., dimensional regularization. Remarkably, this structure allows us to collect our results for the scheme-independent dynamical black-hole Love numbers through ${\cal O} (G^9)$ in terms of \cref{tab}.
\begin{table}
    \centering
    \begin{tabular}{c|c|c|c|c|}
         & $\ZC_\ell$&$\ZC^{(2)}_{\ell}$&$\ZC^{(3)}_{\ell}$&$\ZC^{(4)}_{\ell}$\\
         \vspace{-0.3cm}&&&&\\
         \hline &&&&
         \\
        $l=0$ &$-2$ &$22/3$ &$-628/27$ &$5830/81$ \\&&&&
        \\
        $l=1$ &$-1/6$ &$19/90$ &$-361/2025$ &$-$ \\&&&&
        \\
        $l=2$ &$-1/270$ &$79/28350$ &$-$ &$-$ \\&&&&
        \\
        $l=3$ &$-1/21000$ &$-$ &$-$ &$-$ \\ \hline
    \end{tabular}
    \caption{Table for the scheme-independent entries of $\ZC$.}
    \label{tab}
\end{table}

Moreover, our conjecture enables the resummation of the scheme-independent parts of the Love numbers. 
We find,
\begin{widetext}
\begin{align}
\frac{\bar{\mathcal F_\ell}}{4 \pi \RS^{2\ell+1}} &= 
-\frac{1}{2} \eta \, \ZC_\ell \left(1 + 2\eta\, H(-\eta) + \eta \log\left(\frac{\RS}{R}\right)\right)
+ \frac14 \eta^{2} \, \left(\ZC^{(2)}_{\ell,2}\right)_{0,0} \nn \\
&+ \frac14 \eta^{3} \, \ZC^{(2)}_{\ell} \left(2 H(-\eta) + \log\left(\frac{\RS}{R}\right)\right)
\left(2 + 2\eta\, H(-\eta) + \eta \log\left(\frac{\RS}{R}\right)\right) + \ldots \,,
\end{align}
\end{widetext}
where $\eta = i \omega \RS$ and $H(x)$ is the analytic continuation of the $n$-th harmonic number $H_n$.
The ellipsis represents terms proportional to $\ZC^{(N)}$ with $N>2$, which may also be resummed straightforwardly.
This function has poles at 
\begin{align}
    \omega^*_{ \rm n } =- i \,{\rm n}/\RS \,, \quad {\rm n} \in \mathbb{Z}_{>0}\,,
\end{align}
which coincide with half of the QNM spectrum in the \textit{large-overtone limit}~\cite{Leaver:1985ax,Nollert:1993zz,Andersson_1993,Bachelot:1993dp,Hod:1998vk,Motl:2002hd,Dreyer:2002vy,Motl:2003cd},
\begin{align}
    \omega^\text{QNM}_{\rm n} = -i \, {\rm n } / (2\RS) + \mathcal{O}\left({\rm n}^0\right) \,, \quad 
    {\rm n}\gg1 
    \,.
\end{align}
It would be interesting to promote this analysis to the \textit{$S$-matrix} (see, e.g.,~\cite{Sanchez:1976fcl,Sanchez:1976xm,Sanchez:1977vz,Giddings:2009gj}, and a recent discussion in~\cite{Correia:2025enx}). 
There, unitarity requires that QNMs correspond to simple poles, which will potentially allow us to interpret the higher-order poles we observe here as corrections to the QNM frequencies in the large-overtone limit.
We leave this prospect for future work.

\section{Conclusions}

Shell EFT provides a generic framework for describing a compact body coupled to gravity, it naturally incorporates and leverages known solutions in BHPT in four dimensions, and it leads to vast simplifications in the calculation of Love numbers. 

In the future, it would be interesting to scrutinize the structure of black-hole Love numbers in terms of the Riemann zeta function. 
As a possible extension, we found that the Wilson coefficients $c_{\ell,n}$ also follow the structure of \cref{eq:zeta} for all $\ell$ and $n\geq 1$, where the corresponding matrices are power series in $\RS/R$ with rational coefficients.
Along similar lines, it is natural to ask whether \cref{eq:Love_numbers} may be thought of as a definition of Love numbers at finite $G$.
It would also be interesting to study renormalization in our scheme, especially the interplay between resummation induced by renormalization and the one enabled by our conjecture.

This formalism can next be applied to gravitons and Kerr black holes.
More broadly, the idea of leveraging solutions in BHPT can also be extended to calculations of gravitational self-force~\cite{Mino:1996nk,Quinn:1996am,Detweiler:2000gt,Detweiler:2002mi,Galley:2006gs,Rosenthal:2006iy,Galley:2008ih,Gralla:2008fg,Pound:2009sm,Detweiler:2011tt,Pound:2012nt,Gralla:2012db,Barack:2018yvs} using scattering amplitudes~\cite{Kosmopoulos:2023bwc,Adamo:2023cfp} or worldline methods~\cite{Cheung:2023lnj,Wilson-Gerow:2023syq,Jakobsen:2023tvm,Cheung:2024byb,Akpinar:2025huz,Bjerrum-Bohr:2025bqg}.

\section{Acknowledgments}

We thank M.~Alrashed, V.~Cardoso, V.~De Luca, L.~Dixon, R.~Durrer, S.~Giri, G.~Isabella, A.~Kehagias, P.~Pani, A.~Riotto and F.~Riva for useful discussions and comments on the manuscript.
D.K. is funded by the European Union’s Horizon Europe research and innovation programme under the Marie Skłodowska-Curie grant agreement No. 101208038.
D.P. is supported by the Swiss National Science Foundation under grants no. 200021-205016 and PP00P2-206149.
M.S. is supported by the US Department of Energy under award number DE-SC0024224, the Sloan Foundation, and the Mani L. Bhaumik Institute for Theoretical Physics.

\bibliography{draft}

\clearpage
\newpage
\appendix
\onecolumngrid
\begin{center}
\vspace{0.05in}
 
\vspace{0.05in}
{ \large\it Supplemental Material}
\end{center}
\onecolumngrid
\setcounter{equation}{0}
\setcounter{figure}{0}
\setcounter{section}{0}
\setcounter{table}{0}
\setcounter{page}{1}
\makeatletter
\renewcommand{\theequation}{S\arabic{equation}}
\renewcommand{\thefigure}{S\arabic{figure}}
\renewcommand{\thetable}{S\arabic{table}}

\section{From the Shell to the Worldline}

In this section, we demonstrate the correspondence 
\begin{align}
    \int \frac{\dd^3y \sqrt{\gamma}}{4\pi R^2} \varphi (u^{\mu} \nabla_{\mu})^n \DD \phi \; 
    \rightarrow
    \int \dd\tau \; \nabla_{\langle\ell\rangle}\varphi \, \partial_{\tau}^n\,\nabla_{\langle\ell\rangle}\phi \,,
\end{align}
in the $R\rightarrow 0$ limit, assuming that the limit is well defined, where we normalize a symmetric and trace-free (STF) tensor of $\ell$ indices as 
\begin{align}
    v_{\langle \ell \rangle} = v_{i_1} \ldots v_{i_\ell} + (\text{traces}) \,,
\end{align}
for a vector $\bm v$.
When the limit is ill-defined, the shell acts as a regularization of the worldline. We start by establishing the relation in flat space before we promote it to our curved-space setup. 
We consider coordinates such that $u^\mu=(1,0,0,0)$ and parametrize the worldline in terms of proper time. 

In flat space, i.e., for $G=0$, the identity we wish to prove becomes
\begin{align}
    \left( \partial_{\langle\ell\rangle}\varphi \, \partial_{t}^n\,\partial_{\langle\ell\rangle}\phi \right)\big\rvert_{r=0} = 
    \lim_{R\rightarrow 0} \int \frac{\dd \hat{\bm{n}}}{4\pi} \,\varphi\, \partial_{t}^n \Dd \phi \,,
    \label{eq:flatIdentity}
\end{align}
where $\partial_{\langle\ell\rangle}$ is an STF tensor built out of ordinary spatial derivatives and $\Dd$ is given by \cref{eq:DlDef} with $f(R)=1$. Given that we take the $R\rightarrow 0$ limit and that $\Dd \sim R^{-2\ell}$, we expand both $\phi$ and $\varphi$ to $\mathcal{O}\left(R^{2\ell}\right)$,
\begin{align}
    \phi(t,R,\hat{\bm n}) = \sum_{a=0}^{2\ell} \frac{R^a}{a!} \left(\hat{\bm n}\cdot \bm \partial\right)^a \phi \big\rvert_{r=0} \,, \qquad
    \varphi(t,R,\hat{\bm n}) = \sum_{b=0}^{2\ell} \frac{R^b}{b!} \left(\hat{\bm n}\cdot \bm \partial\right)^b \varphi \big\rvert_{r=0} \,.
\end{align}
We use the identity 
\begin{align}
    (\hat{\bm n}\cdot\hat{\bm z})^a = \hat{n}_{\langle a\rangle}\hat{z}_{\langle a\rangle} \, + \text{lower}\,,
\end{align}
where `lower' refers to terms of the form $\hat{n}_{\langle b\rangle}\hat{z}_{\langle b\rangle}$ with $b<a$, to write
\begin{align}
    \left(\hat{\bm n}\cdot \bm \partial\right)^a \phi\big\rvert_{r=0} = \hat{n}_{\langle a\rangle} \partial_{\langle a\rangle} \phi\big\rvert_{r=0} \, + \text{lower}\,.
\end{align}
We collect all terms proportional to $\hat{n}_{\langle a\rangle}$,
\begin{align}
    \phi(t,R,\hat{\bm n}) = \sum_{a=0}^{2\ell} \frac{R^a}{a!}  \left( \hat{n}_{\langle a\rangle} \partial_{\langle a\rangle}\phi \big\rvert_{r=0} + \text{lower} \right) =\sum_{a=0}^{2\ell} \frac{R^a}{a!} \hat{n}_{\langle a\rangle}  \Big(\partial_{\langle a\rangle}\phi \big\rvert_{r=0} + \mathcal{O}\left(R\right) \Big)\,,
\end{align}
where the $\mathcal{O}\left(R\right)$ terms come from the `lower' contributions of higher $a$.
With these we can compute $\Dd \phi$ as 
\begin{align}
    \Dd \phi &= \frac{(-1)^\ell (2\ell+1)}{2^\ell R^{2\ell}} \sum_{a=\ell}^{2\ell} \frac{R^a}{a!}
     E_{\ell a} \hat{n}_{\langle a\rangle} \Big( \partial_{\langle a\rangle} \phi\big\rvert_{r=0} + \mathcal{O}\left(R\right) \Big)\,,
     \label{eq:expansionPhi}
\end{align}
where we used the fact that $E_{\ell a}$ (defined in \cref{eq:E_projector}) vanishes if $a<\ell$, together with the identity
\begin{align}
    \nabla_\Omega^2 \, \hat{n}_{\langle \ell\rangle} = -\ell(\ell+1)\, \hat{n}_{\langle \ell\rangle} \,,
\end{align}
which itself follows from 
\begin{equation}
    P_{\ell}(\hat n\cdot \hat z) = \frac{2\ell!}{2^{\ell}(\ell!)^2}\hat n_{\langle\ell\rangle } \hat z_{\langle \ell \rangle }\,.
    \label{eq:LegendreDef}
\end{equation}
Next, we similarly rearrange $\varphi$,
\begin{align}
    \varphi(t,R,\hat{\bm n}) = \sum_{b=0}^{2\ell} \frac{R^b}{b!}  \left( \hat{n}_{\langle b\rangle} \partial_{\langle b\rangle}\varphi \big\rvert_{r=0} + \text{lower} \right) =\sum_{b=0}^{2\ell} \frac{R^b}{b!} \hat{n}_{\langle b\rangle}  \Big(\partial_{\langle b\rangle}\varphi \big\rvert_{r=0} + \mathcal{O}\left(R\right) \Big)\,.
     \label{eq:expansionVarPhi}
\end{align}
Finally, using \cref{eq:expansionPhi,eq:expansionVarPhi} in the right-hand side of \cref{eq:flatIdentity}, together with 
\begin{equation}
    \mathds{1}_{\langle a\rangle \langle b\rangle}=(2a+1)\frac{(2a)!}{2^{a}(a!)^2}
    \int\frac{\dd \hat{\bm{n}}}{4\pi}\,\sv{n}{a}\sv{n}{b} \,,
    \label{eq:id_tensor}
\end{equation}
we arrive at the left-hand side.

In our shell geometry, the unit vector orthogonal to the shell takes the form $\hat n_\mu = (0,\sqrt{f(R)},0,0)$.
Tracing through the above derivation, we see that to lift our identity in \cref{eq:flatIdentity} to curved space, i.e., for $G \neq 0$, we need to promote $\Dd \rightarrow \DD$.
This establishes the correspondence between the shell and worldline actions.

\section{Results for Love numbers}

In this section we list the complete results for the black-hole Love numbers $\bar{\mathcal{F}}_\ell$ up to order $\mathcal{O}\left(G^9\right)$ and discuss the comparison to other schemes. 
The scheme-independent parts of these results are also given in \cref{tab}.
The corresponding results for the Wilson coefficients are given in the ancillary files.
\begin{align}
    \bar{\mathcal{F}}_0&= 
    4 i \pi  \omega  \RS^2 -\frac{1}{2} \pi  \omega ^2 \RS^3 \left(8 \log \left(\frac{\RS}{R}\right)+11\right) +\frac{1}{6} i \pi  \omega ^3 \RS^4 \left(-88 \log \left(\frac{\RS}{R}\right)+8 \pi ^2-25\right)\nn\\
    & -\frac{\pi  \omega ^4 \RS^5}{8640} \left(69120 \zeta_{3}-63360 \log ^2\left(\frac{\RS}{R}\right)-221280 \log \left(\frac{\RS}{R}\right)+42240 \pi ^2-7583\right)\nn\\
&-\frac{i \pi  \omega ^5 \RS^6}{4320} \left(384 \pi ^4-150720 \log ^2\left(\frac{\RS}{R}\right)+21120 \pi ^2 \log \left(\frac{\RS}{R}\right)-165232 \log \left(\frac{\RS}{R}\right)+126720 \zeta_{3} \right. \nn \\
& \qquad \qquad +36880 \pi ^2+71417\bigg)\nn\\
&-\frac{\pi  \omega ^6 \RS^7}{21772800} \left( -174182400 \zeta_{5} +10644480 \pi^4+253209600 \log ^3\left(\frac{\RS}{R}\right)+1911047040 \log ^2\left(\frac{\RS}{R}\right)\nn\right.\\
& \qquad \qquad-506419200 \pi ^2 \log \left(\frac{\RS}{R}\right)+1101382800 \log \left(\frac{\RS}{R}\right)-878169600 \zeta_{3}\nn\\
&\qquad \qquad\left.-638668800 \zeta_{3} \log \left(\frac{\RS}{R}\right) -277589760 \pi ^2-936497323\right)\nn\\
&+\frac{i \pi  \omega ^7 \RS^8 }{1360800}\left(11520 \pi ^6+39916800 \zeta_{5}-97944000 \log ^3\left(\frac{\RS}{R}\right)+15825600 \pi ^2 \log ^2\left(\frac{\RS}{R}\right) \right.\nn\\
&\qquad \qquad -4336080 \pi ^4 -264464760 \log ^2\left(\frac{\RS}{R}\right)+443520 \pi ^4 \log \left(\frac{\RS}{R}\right)+79626960 \pi ^2 \log \left(\frac{\RS}{R}\right)\nn\\
&\qquad \qquad -13305600 \pi ^2 \zeta_{3}+49764960 \zeta_{3}+189907200 \zeta_{3} \log \left(\frac{\RS}{R}\right)\nn\\
&\qquad \qquad \left. +25605367 \log \left(\frac{\RS}{R}\right)+22945475 \pi^2+58628869\right)\nn\\
&+\frac{\pi  \omega ^8 \RS^9}{438939648000} \left(-3511517184000 \zeta_{7}-30628233216000 \zeta_{3} \log ^2\left(\frac{\RS}{R}\right)-12875563008000 \zeta_{5} \log \left(\frac{\RS}{R}\right)\right.\nn\\
&\qquad \qquad -136581761433600 \zeta_{3} \log \left(\frac{\RS}{R}\right)+7898204160000 \log ^4\left(\frac{\RS}{R}\right)+107694565171200 \log ^3\left(\frac{\RS}{R}\right)\nn\\
&\qquad \qquad -31592816640000 \pi ^2 \log ^2\left(\frac{\RS}{R}\right)+164490996579840 \log ^2\left(\frac{\RS}{R}\right)\nn\\
&\qquad \qquad  +1020941107200 \pi ^4 \log \left(\frac{\RS}{R}\right) -56870501990400 \pi ^2 \log \left(\frac{\RS}{R}\right)+3712941711360 \pi ^4\nn\\
&\qquad \qquad -32042594304000 \zeta_{5} +12875563008000 \zeta_{3}^2+20418822144000 \pi ^2 \zeta_{3}\nn\\
&\qquad \qquad -24046102886400 \zeta_{3}+34062336000 \pi ^6 +2753089059840 \pi ^2\nn\\
&\qquad \qquad \left.  -123148321363200 \log \left(\frac{\RS}{R}\right)+2165341072007 \right) +\mathcal{O}(\RS^{10})
\,,
\end{align}

\begin{align}
\bar{\mathcal{F}}_{1} &=
\frac{1}{3} i \pi  \omega  \RS^4-\frac{1}{2880}\pi  \omega ^2 \RS^5 \left(960 \log \left(\frac{\RS}{R}\right)+199\right)+\frac{1}{360} i \pi  \omega ^3 \RS^6 \left(-152 \log \left(\frac{\RS}{R}\right)+40 \pi ^2+383\right)\nn\\
&-\frac{\pi  \omega ^4 \RS^7}{36288000} \left(24192000 \zeta_{3}-7660800 \log ^2\left(\frac{\RS}{R}\right)+33754560 \log \left(\frac{\RS}{R}\right)+5107200 \pi ^2+15646349\right)\nn\\
&-\frac{i \pi  \omega ^5 \RS^8}{2268000} \left(+16800 \pi ^4-606480 \log ^2\left(\frac{\RS}{R}\right)+319200 \pi ^2 \log \left(\frac{\RS}{R}\right)+1915200 \zeta_{3}-703220 \pi ^2\right.\nn\\
&\qquad \qquad \left.+3680636 \log \left(\frac{\RS}{R}\right)-4322939\right)\nn\\
&-\frac{\pi  \omega ^6 \RS^9 }{3657830400000}\left(-2438553600000 \zeta_{5}-3088834560000 \zeta_{3} \log \left(\frac{\RS}{R}\right)+326043648000 \log 
^3\left(\frac{\RS}{R}\right)\right.\nn\\
&\qquad\qquad -2658323404800 \log ^2\left(\frac{\RS}{R}\right)-652087296000 \pi ^2 \log
\left(\frac{\RS}{R}\right)+7783050240000 \zeta_{3}\nn\\
&\qquad \qquad \left.+4705683548160 \log \left(\frac{\RS}{R}\right)+51480576000
\pi ^4+1978709913600 \pi ^2+6150318386659\right)+\mathcal{O}(\RS^{10})
\,,
\end{align}

\begin{align}
    \bar{\mathcal{F}}_{2}&=
    \frac{1}{135} i \pi  \omega  \RS^6-\frac{\pi  \omega ^2 \RS^7}{10886400} \left(80640 \log \left(\frac{\RS}{R}\right)+13663\right)+\frac{i \pi  \omega ^3 \RS^8}{340200} \left(-1896 \log \left(\frac{\RS}{R}\right)+840 \pi ^2+8387\right)\nn\\
    &-\frac{\pi  \omega ^4 \RS^9}{15362887680000} \left(227598336000 \zeta_{3}-42810163200 \log ^2\left(\frac{\RS}{R}\right)+28540108800 \pi ^2\right.\nn\\
    &\qquad\qquad \left.+362638725120 \log \left(\frac{\RS}{R}\right)-69269259397\right)+\mathcal{O}(\RS^{10})
    \,,\\
    \bar{\mathcal{F}}_{3}&=
    \frac{i \pi  \omega  \RS^8}{10500}-\frac{\pi  \omega ^2 \RS^9 }{135475200000}\left(12902400 \log \left(\frac{\RS}{R}\right)+1864267\right)+\mathcal{O}(\RS^{10})
    \,,\\
    \bar{\mathcal{F}}_4&=\mathcal{O}(\RS^{10}) \,.
\end{align}

To identify the scheme-independent pieces of our results, we perform the shift
\begin{align}
    \log\left(\RS/R\right) \, \rightarrow \, \log\left(\RS/R\right) \, + \, Q\,,
\end{align}
for some rational number $Q$ and organize the result in terms of transcendental weight. Terms unaffected by this shift are scheme-independent. For instance, in
\begin{align}
    \bar{\mathcal{F}}_0&= \ldots 
    -\frac{\pi  \omega ^4 \RS^5}{8640} \left(69120 \zeta_{3}-63360 \log ^2\left(\frac{\RS}{R}\right)-221280 \log \left(\frac{\RS}{R}\right)+42240 \pi ^2-7583\right)
    + \ldots \,,
\end{align}
the coefficients of $\zeta_3$, $\log^2(\RS/R)$ and $\pi^2$ are scheme independent, while that of $\log (\RS/R)$ and the rational number are not.
The scheme-independent pieces of our results for the black-hole Love numbers $\bar{\mathcal{F}}_\ell$ and those computed in dimensional regularization through $\mathcal{O}\left(G^7\right)$~\cite{Caron-Huot:2025tlq} agree upon the identification $\mu=1/R$.
To facilitate comparison, we note that our $c_{\ell,n}$ and $\omega$ should be compared to $(-1)^n C_{\ell,n}$ and $-\omega$ of~\cite{Caron-Huot:2025tlq} due to the different placement of the time derivatives in the action and the different convention in, e.g., \cref{eq:ansatz-out}; these two signs cancel in the definition of the Love numbers $\bar{\mathcal{F}}_\ell$ (see \cref{eq:wilson_coeff}).

\end{document}